\documentclass [12pt]{article}
\usepackage[utf8]{inputenc}
\usepackage{amsmath,graphicx}
\usepackage{geometry}
\usepackage{authblk}
\usepackage{xcolor}
\usepackage{bbold}
\usepackage{soul,comment}\usepackage{tikz,xcolor,hyperref}

\providecommand{\keywords}[1]
{
  \small	
  \textbf{\textit{Keywords---}} #1
}

\definecolor{lime}{HTML}{A6CE39}
\DeclareRobustCommand{\orcidicon}{%
	\begin{tikzpicture}
	\draw[lime, fill=lime] (0,0) 
	circle [radius=0.16] 
	node[white] {{\fontfamily{qag}\selectfont \tiny ID}};
	\draw[white, fill=white] (-0.0625,0.095) 
	circle [radius=0.007];
	\end{tikzpicture}
	\hspace{-2mm}
}

\foreach \x in {A, ..., Z}{%
	\expandafter\xdef\csname orcid\x\endcsname{\noexpand\href{https://orcid.org/\csname orcidauthor\x\endcsname}{\noexpand\orcidicon}}
}

\usepackage{mathabx, MnSymbol, wasysym}
\usepackage{setspace}
 \makeatletter
\def\@fnsymbol#1{\ensuremath{\ifcase#1\or \dagger \or \star\or
   \mathsection\or \mathparagraph\or \|\or **\or \dagger\dagger
   \or \ddagger\ddagger \else\@ctrerr\fi}}
    \makeatother
\newcommand{\bea}{\begin{eqnarray}}
\newcommand{\eea}{\end{eqnarray}}
\newcommand{\ang}{\mbox{\normalfont\AA}}

\title{Structural Deformation and Metal-Semiconductor Transition in Coupled Carbon Chains}

\date{}

 \author[1]{Rudranil Basu \thanks{rudranilb@goa.bits-pilani.ac.in}\orcidA{}
 }
\author[2]{Swastibrata Bhattacharyya \thanks{Corresponding author. Tel: +91-832-2580-365. E-mail: swastibratab@goa.bits-pilani.ac.in (Swastibrata Bhattacharyya)}\orcidB{}}

\affil[1,2] {Department of Physics, Birla Institute of Technology and Science Pilani, Zuarinagar, Goa 403726, India}
\affil[1]{Center for Fundamental Laws of Nature, Harvard University, Cambridge,
MA 02138, USA}

\begin{document}
\maketitle
\begin{abstract}
The transition between gapped (semiconducting) and gapless (metallic) phases and tunability of bandgap in materials is a very lucrative yet considerably challenging goal for new-age device preparation. For bulk materials and for two-dimensional layered systems, this is a rapidly expanding field. We theoretically propose a one-dimensional pure carbon material with a tunable bandgap. We find that two parallel coupled polyyne chains show metallic behaviour with bands crossing on the Fermi level, unlike the single semiconducting chain. The number of nodal points (two) is robust under transverse and longitudinal strain, indicating the symmetry-protected nature of the metallic phase. Sliding one chain with respect to the other breaks reflection symmetry and a clear bandgap opens up at the nodes, leading to a gapped phase. By varying the slide parameter, the bandgap can be tuned efficiently. This work initiates and indicates possible topological phases of real one-dimensional materials without the involvement of edge modes.
\end{abstract}

\section{Introduction}
 Tuning of electronic properties and metal to insulator transition in low dimensional materials is extremely important from a materials engineering perspective. Such transitions and/or tuning of bandgap in bulk and 2D layered materials can be obtained by
changing the materials chemistry; application of external electric and magnetic fields and introducing structural deformation such as  application of strain and defect engineering \cite{Lei2018,Wang1994,PYoshizawa1995,BAFEKRY2020371,Tiwari2002}. Apart from technological advancement in materials designing \cite{kezilebieke2020}, these methods of bandgap tuning and metal/ semi-metal to insulator transition lend a deeper theoretical understanding of many-body low energy physics. Among these methods, structural deformation such as application of strain and change in stacking is particularly very important because of the practical feasibility and observation of interesting physics induced by it in some materials. Structural deformations have been observed to be the cause of various electronic phase transitions including topologically non-trivial and trivial phase transitions involving semi-metal to insulator \cite{Lin2020,petrov2017}.  The topologically non-trivial phase of matter was first observed in the 2D avatar of pure carbon, ie. graphene \cite{Novoselov2004} which shows linear dispersion. The present work is about searching for metal-insulator phase transition in one-dimensional pure carbon device, inspired by a series of curious results in bilayer graphene found in the past decade. Whereas flat-bands appear in twisted bilayer graphene at magic angles \cite{cao2018unconventional,tarnopolsky2019origin}, Lifshitz phase transition appears for sliding \cite{son2011electronic,bhattacharyya2016lifshitz}, where the system retains its semi-metallic feature. Hence it is imperative that one searches for such transitions in 1D pure carbon material.

 One-dimensional materials such as nanotubes, nanoribbons, nanorods, and nanowires posses various interesting transport and electronic properties \cite{Chen2021,Bachtold2000,Ezawa2006}. These properties can be also explored under the application of structural deformation \cite{Cretu2013,lin2020visualization} for their potential applications in miniaturised devices. Among these 1D forms of materials, linear atomic chains are the thinnest 1D material and the interest for this work.
  
  Among the 1D allotropes of carbon - cumulene and polyyne are the two forms in the class of atomic chains. Polyyne, having alternating single and triple bonds is more stable as per the Peierls' theorem\cite{peierls1991more}.   
Polyyne chains are found in biological entities\cite{Shi2006}, astronomical objects\cite{McCarthy2001} and as intermediates for organic synthesis\cite{Kroto1993,Richter2000}. Experimentally, long polyyne chains have been synthesised\cite{Cataldo2004,Kutrovskaya2020,Casillas2014} with interesting optical\cite{Kutrovskaya2020} and mechanical properties\cite{Nair2011}.  
The single polyyne chain is  semiconducting \cite{lambropoulos2017electronic, al2014electronic, kartoon2018driving} and bandgap tuning has been observed under strain\cite{Li_2018,Cretu2013}. 
Also (semi-conducting) polyyne to (metallic) cumulene transition has been demonstrated experimentally under application of strain \cite{la2015strain}. In the theoretical front, the effect of coupling on finite length carbon chains bridging graphene electrodes on conductivity has recently been studied, \cite{liang2019influence} which shows noticeable variation depending upon even or odd number of atoms in the chain.

Save a few quasi 1D topologically non-trivial systems \cite{lin2020visualization,lutchyn2010majorana} it is rare to observe metal-insulator transition in one-dimensional systems. In this work, motivated by the exotic phases of bilayer graphene, we have investigated the electronic properties of coupled polyyne chains. We show, using the tight-binding model and \textit{ab-initio} density functional theory (DFT), that when two chains are brought close, the lowest energy bands cross each other at the Fermi level with linear dispersion. The coupled chain system is interesting because it shows the transition from metallic to insulator phase under a change in stacking pattern by sliding. In fact, the phenomenon of bandgap opening under a change in stacking is very similar to the one observed in materials with high spin-orbit coupling (SOC), reminiscent of band inversion. 

\section{Tight binding formalism for AA stacked coupled chains}

\subsection{The metallic phase}

The carbon atoms in a polyyne chain are under $sp$ hybridization and form alternating single and triple bonds with the nearest neighbour atoms to construct a linear chain of the form $\mathrm{(-C \equiv C-)_n}$. There are two $\pi$ electrons per carbon atom in this linear 1D structure. The unit cell of polyyne consists of two carbon atoms marked as sublattice $a$ ($\Tilde{a}$) and $b$ ($\Tilde{b}$) as shown in Fig. \ref{fig1}(a,b). The structure of the system of coupled polyyne chains placed parallel to each other, as in Fig. \ref{fig1}(b), such that the single bonds of one chain are aligned to their counterparts of the other one, will be named as `AA' stacking. Assuming only nearest neighbour (NN) hopping of electrons, the intra-chain and the inter-chain hopping energy costs are respectively $t_1, t_2$ and $\gamma_1$.

In the tight-binding approximation, the total Hamiltonian for this system is
\bea \label{ham_double_chain}
H&=&t_{1}\sum _{p}a^{\dagger}_{p} b_{p}+t_{2}\sum_{p} a_{p+1}^{\dagger}b_{p} + t_{1}\sum _{p}\Tilde{a}^{\dagger}_{p} \Tilde{b}_{p}+t_{2}\sum_{p} \Tilde{a}_{p+1}^{\dagger}\Tilde{b}_{p} \nonumber \\
&+& \gamma_{1}\sum _{p}\left( \Tilde{a}^{\dagger}_{p} a_{p} + \Tilde{b}^{\dagger}_{p} b_{p} \right) + \mbox{ h.c.}
\eea
Going to the Fourier space, we have:
\bea \label{double_chain_fourier}
H&=& \sum_{k} \Psi^{\dagger}_k\begin{pmatrix} 0 & f_k & \gamma_1 & 0\\ f^{\star}_k & 0 &0 & \gamma_1\\ \gamma_1 & 0 &0 & f_k \\ 0 & \gamma_1 & f^{\star}_k & 0
\end{pmatrix} \Psi_{k}\\
\mbox{with }~\Psi^{\dagger}_k &=&\begin{pmatrix} c^{\dagger}_k & d^{\dagger}_k & \Tilde{c}^{\dagger}_k & \Tilde{d}^{\dagger}_k \end{pmatrix}, \nonumber
\eea
where $c_k$'s are the Fourier basis modes corresponding to the real space modes $a_p$'s and $f_k =t_1 e^{ik \delta_1} + t_2 e^{-ik \delta_2} $. The characteristic polynomial of the above matrix is bi-quadratic and hence can be trivially diagonalized to give the following 4 bands:
\bea \label{2_chain_BS}
\mathcal{E}^{(m,n)}_k 
= (-1)^m \left( - \gamma_1 + (-1)^n \sqrt{t^2_1 + t^2_2 + 2 t_1 t_2 \cos(kA)}  \right)
\eea
 The bands $\mathcal{E}^{(1,1)}_k$ and $\mathcal{E}^{(2,1)}_k$, respectively the valence band (VB) and the conduction band (CB) give rise to a couple of crossing nodes at Fermi level, located at momenta $k_1 = \theta /A$ and $k_2 = (2 \pi- \theta)/A$, where $\theta = \arccos\left(\dfrac{\gamma_1^2 -t^2_1- t^2_2}{2 t_1 t_2}\right)$ as long as $\gamma_1$ is restricted as $|\gamma_1| \in \left[|t_1-t_2|, |t_1+t_2|\right]$. Hence the metallic phase is observed for this range of $\gamma_1$.
 Expanding the dispersion relation $\mathcal{E}^{(2,1)}_q$ around $k_1 = \theta /A$, we find, up to  the linear term:
\bea
\mathcal{E}^{(2,1)}_k = \frac{t_1 t_2 \sin{\theta}}{\gamma_1} A\,q .
\eea
where $q = k-k_1$. This implies that the Fermi velocity, $v_F = \frac{t_1 t_2 \sin{\theta}}{\hbar \gamma_1} A.$

About one of the points $k = k_1$, the above Hamiltonian's kernel in equation\eqref{double_chain_fourier} has the following expected low energy behaviour:
\bea \label{low energy}
\mathcal{H}_q&=& \begin{pmatrix} ( \Vec{a} + q\,\Vec{b}) \cdot \Vec{\sigma} & \gamma_1 \mathbb{1}_2 \\ \gamma_1 \mathbb{1}_2 & ( \Vec{a} + q\,\Vec{b}) \cdot \Vec{\sigma}
\end{pmatrix},  \\
 && \vec{a} = (\Re f(k_1), - \Im f(k_1),0),\, \vec{b} = (\Re  f'(k_1), - \Im f'(k_1),0) \nonumber .
\eea
This acts on the 4-component fermion $\Psi_{q+k_1}$. We note that the Hamiltonian in \eqref{double_chain_fourier} enjoys time reversal, charge conjugation and chiral symmetry and falls in the BDI topological class \cite{kitaev2009periodic}.

The intra-chain interactions between sites are due to bond formation, whereas the inter-chain ones are effectively Van der Waals. 
With the increase in the inter-chain distance $d$, the hopping parameter, $\gamma_1$ falls fast towards zero. As the strength $|\gamma_1|$ reaches the value $|t_1-t_2|$ from above, the two nodal points merge, giving rise to parabolic dispersion at the Fermi level with the CB and the VB touching each other. Further increase in separation results in disappearing inter-chain interaction and in opening up of a gap, making it semiconducting as a single polyyne chain\cite{lambropoulos2017electronic}.

\subsection{Estimating the tight-binding Parameters from Density Functional Theory}

\begin{figure}[hbt!]
    \centering
    \includegraphics[width=0.9\columnwidth]{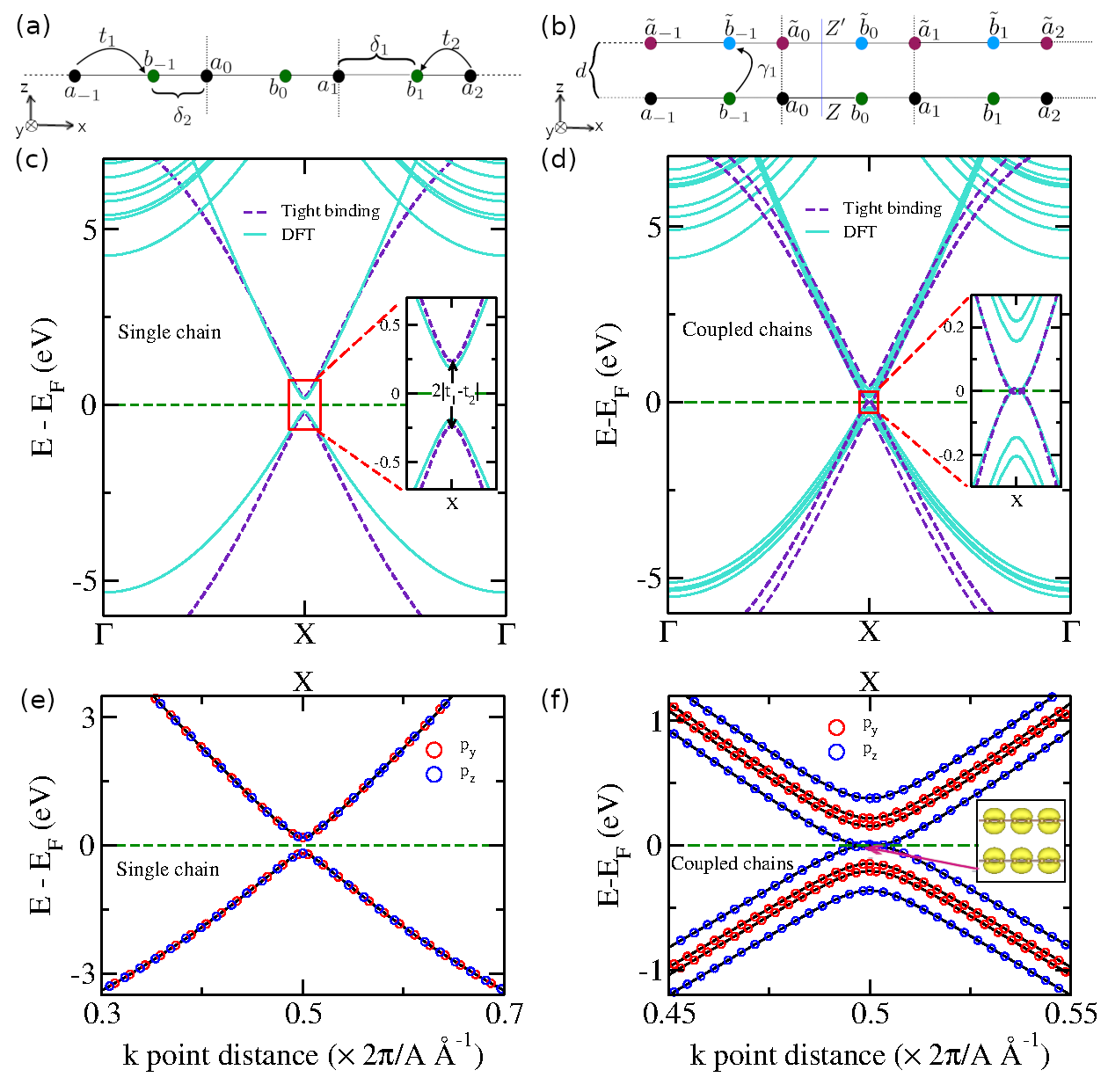}
    \caption{ \textbf{Lattice structures, tight binding definitions, and band structures of single and coupled Polyyne chain. } (\textbf{a}) Hopping terms between the shorter and the longer bonds in polyyne. C atoms at $a$ and $b$ sites are represented by black and green circles. It should be noted that this is a cartoon diagram of the lattice structure, meant to show that $\delta_1 > \delta_2 $ and the lengths depicted are not to be scaled. In reality, $\delta_2 \sim 0.97 \delta_1 $. (\textbf{b}) Two close-by polyyne chains with inter-chain hopping $\gamma_1$ shown in AA configuration. C atoms at different chains and atomic sites are denoted by different colors. The system evidently has reflection (parity) symmetry about the line $ZZ'$ or any other line produced by lattice translation $A$. Equivalently there is one more line of parity symmetry (not shown in the figure) for each unit cell. The vertical dotted lines represent the unit cell for both (\textbf{a}) and (\textbf{b}).
    Bandstructure of (\textbf{c})  single and (\textbf{d}) coupled polyyne chain for relaxed inter-chain separation and for AA stacking. The dotted blue line represents tight-binding and solid turquoise line represents DFT bandstructures. The orbital projected bandstructures are plotted for (e) single and (f) coupled chains  near the X point of the Brillouin zone. The contribution of various orbitals to the bands are denoted by different colours. The Fermi level is shown with green dotted line. {The band decomposed charge density of the highest occupied molecular orbital (HOMO) at the X point is shown in the inset of (f). Same axes direction as in (a) has been followed for the inset plot.}}
    \label{fig1}
\end{figure}

 Since the bandstructure in equation\eqref{2_chain_BS} are sensitive to the values of the tight-binding parameters, estimating them via first-principles calculations (DFT) is important to predict the phases of the electronic structure. The bandstructure for the single and coupled polyyne chains as calculated using DFT is shown in Figure. \ref{fig1}(c) and (d), respectively. To gain further insights into the bandstructure, the corresponding orbital projected bandstructures were calculated and plotted in Figure. \ref{fig1}(e) and (f), for the single and the coupled chains, respectively. Details of the DFT calculations are presented in the subsection \ref{appcomp}. The single-chain shows parabolic dispersion near the Fermi level both in the VB and CB.  These low energy bands are originated from the two pairs of $\pi$ electrons at $p_y$ and $p_z$ orbitals of the two C atoms in the unit cell, as seen in Figure. \ref{fig1}(e). Due to the rotational symmetry of the single-chain about the x-axis, these $\pi$ electrons are in doubly degenerate states causing complete overlap of bands emanating from the $p_y$ and $p_z$ orbitals; both for the CB as well as the VB. A bandgap of 0.36 eV is observed at the X-point of the Brillouin zone.
 
 When a second polyyne chain is placed near the first chain, separated along the z-direction, the rotational symmetry about the x-axis is broken. As expected, this symmetry breaking lifts the above degeneracies of the single-chain states, causing a split between the $p_y$ and $p_z$ bands.
 Thus, there are four $p_z$ bands, two (from two chains) each in the CB and in the VB (Figure. \ref{fig1}(f)). 
 The splitting is stronger in the $p_z$ bands compared to the $p_y$ bands in this structural configuration. Therefore, the remaining two $p_y$ bands lie in between these  $p_z$ bands in both CB and the VB (Figure. \ref{fig1}(d,f)). {Thus for the coupled chain, the bands closest to the Fermi level are mostly influenced by the pi ($p_z$) orbitals, which are in the plane formed by the two chains (xz plane in our geometry). The $p_y$ orbitals, which are orthogonal to this plane, do not contribute to these low energy states.}
 {The band decomposed charge density plots for the highest occupied molecular orbital at the X point (Figure. \ref{fig1}(f) inset) clearly shows the contribution from the $p_z$ orbital to the VBM.} The bandstructure of the coupled chains shows a slight overlap between the CB and the VB at the Fermi level, indicating it to be semimetal.

 To obtain the tight-binding parameters from the DFT bandstructure, we fitted for the VB\footnote{In fact, near the Fermi level, the system has approximate particle-hole symmetry. Hence, fitting for the tight-binding parameters is the same for both CB and VB, to the significant digits.}. This resulted in a good fit for the two bands closest to the Fermi level as shown in Figure. \ref{fig1}(c) and (d). The fitted parameters obtained for a single polyyne chain are: $t_1 = 3.682 \, \mathrm{eV}$ and $ t_2 = 3.92 \, \mathrm{eV}$. We here note a slight departure in hopping parameters from that quoted in\cite{al2014electronic} for a single polyyne chain. The inter-chain hopping parameter has been estimated to have the value $\gamma_1 = 0.247\, \mathrm{eV}$ from the two polyyne chains in the relaxed configuration, ie. separated by $3.78 \, \ang$. 

\subsection{Computational details} \label{appcomp}
All the \textit{ab-initio} density functional theory (DFT) calculations were performed using the scientific package, Vienna $ Ab-initio$ Simulation Package (VASP)\cite{Kresse1993}. All-electron projector augmented wave (PAW) method\cite{Blochl94,Kresse99} was used to describe the interaction between ions and electrons and the Perdew-Burke-Ernzerhof (PBE)\cite{Kresse99} generalized gradient approximation (GGA) was used to account for the electronic exchange and correlation. The optimization of the lattice parameter for the single polyyne chain was done using the conjugate gradient algorithm with a minimum force cutoff criteria of 0.0001 eV/\AA ~on every atom and an energy cutoff for the plane-wave basis set as 400 eV.  Since PBE-GGA can not properly describe Peierls' distortion and hence the bond length alteration (BLA) \cite{Wanko2016, Yang2006}, the ratio of the bond lengths was kept fixed to prevent the structure from relaxing into that of the cumulene one. 
{To check if there is any effect on the intra-chain bond lengths due to the structural changes caused by sliding, we performed a few calculations using LDA functional by allowing the atoms to relax under sliding and at an interlayer distance of 3.55 \AA. No noticeable change in the bond length was observed. Therefore, we believe that our approximation of no change in bond length under strain and sliding is reasonable.}

The unit cells used for the calculations are shown in Fig. \ref{fig1} (a) and (b) for the single chain and the coupled chains, respectively. The direction of the chains is considered periodic along the x-axis and for the coupled chain, the second chain is placed at a distance $d$ along the z-axis. A sufficient vacuum was used in both directions perpendicular to the chain length (i.e., along y- and z-axis) to avoid any interaction between the  periodic images. A well-converged Monkhorst-Pack k-point set of 11$\times$1$\times$1 was used for the structure optimization. An optimized lattice parameter of 2.56 \AA~and bond lengths of 1.298 \AA~and 1.262 \AA~were obtained for the polyyne chain that agrees well with the previously reported DFT-LDA values\cite{rusznyak2005bond,al2014electronic}. To obtain the relaxed inter-chain distance for the coupled polyyne chain, optB88-vdW functional\cite{Klime2009,Klime2011} as implemented in VASP was used to treat the week van der Waals (vdW) interaction holding the two chains. The relaxed inter-chain distance calculated was 3.78 \AA.

\section{Structural Deformations}
We will now consider three ways of deforming the structure of the coupled polyyne chains and study the response of the bands close to the Fermi level under these deformations. The number of nodal points at the Fermi level is robust under strains, but not under sliding, which opens up a bandgap. 

\subsection{Transverse strain}
We first applied transverse compressive strain ($\varepsilon_{\mathrm{Tr}}$) along the $z$ direction to study the effect of varying inter-chain distance on the dispersion of the low energy bands in the coupled polyyne chains. To incorporate this effect in our tight-binding model of equation\eqref{double_chain_fourier} we need to introduce in the inter-chain hopping parameter $\gamma_1$, a functional dependence on the inter-chain separation, $d$. As $d$ decreases, the inter-chain interaction increases , and therefore the absolute value of $\gamma_1$ should increase. Motivated by similar modelling applied for 2D bilayers\cite{kariyado2019flat,fang2016electronic,padhi2019pressure}, we assume a gaussian dependence:
\bea \label{par}
\gamma_1 (d) = t_0 \exp\left(- \frac{d^2}{\kappa \, d^2_0} \right). \eea
Here $d_0 = 3.78 \, \ang$ is the relaxed inter-chain separation and $ \kappa = 0.4$  is the gaussian width controlling parameter. Along with these, choosing  $t_0 = 3.063 \, \mathrm{eV}$, the analytical forms of the couple of bands closest to the Fermi level given by equation\eqref{2_chain_BS} gives the fitting with the DFT results.

\begin{figure}[h]
    \centering
    \includegraphics[width=\columnwidth]{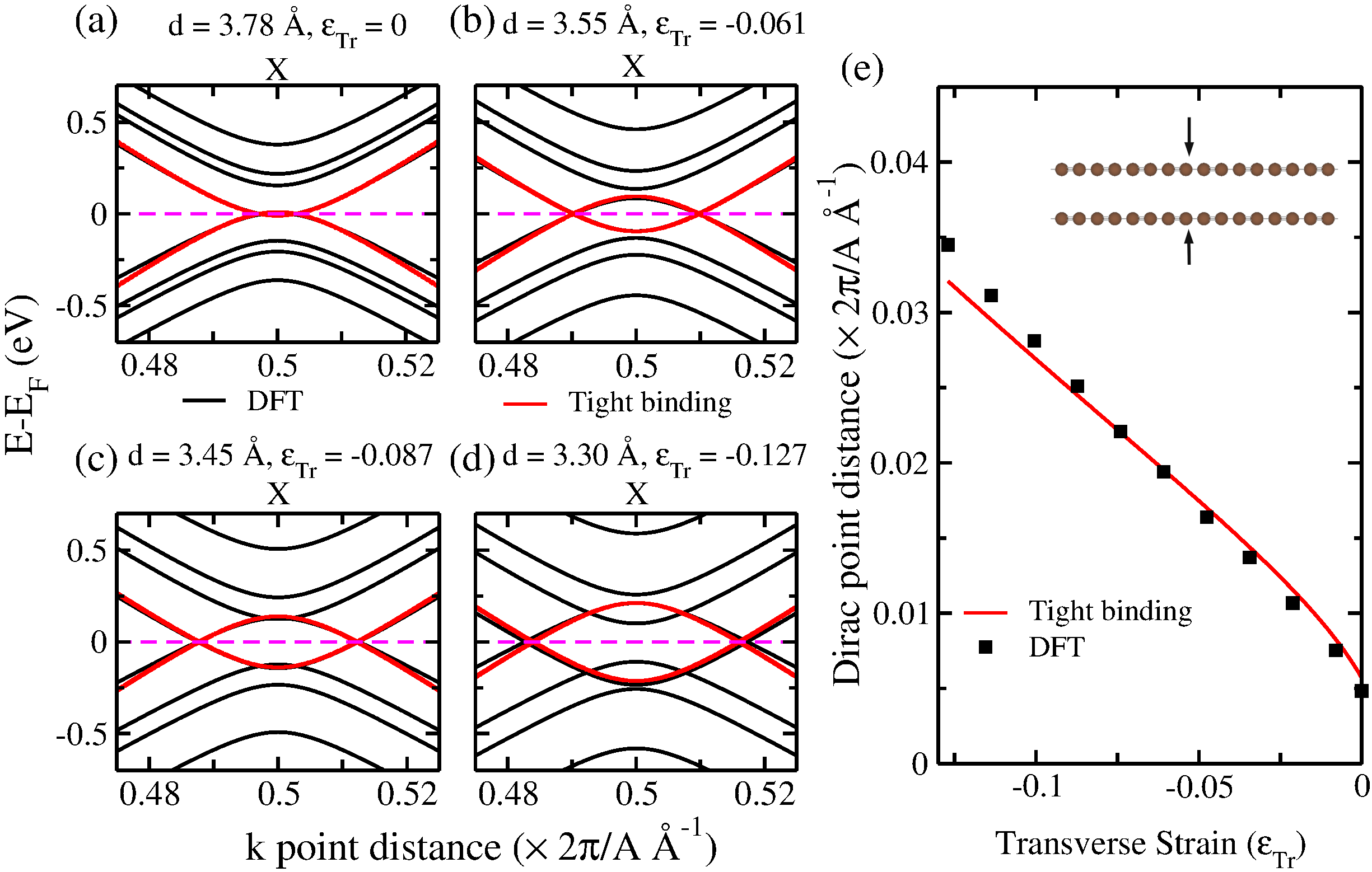}
    \caption{\textbf{Effect of applied transverse strain on bandstructure, calculated using tight-binding and DFT for the coupled chains .} The energy dispersion of the coupled polyyne chain for the lowest energy bands with increasing transverse strain. Bandstructure of coupled polyyne chain under strain (\textbf{a}) $\varepsilon_{\mathrm{Tr}} = 0$, (\textbf{b}) $\varepsilon_{\mathrm{Tr}} = -0.061 $, (\textbf{c}) $\varepsilon_{\mathrm{Tr}} = -0.087 $ and (\textbf{d}) $\varepsilon = -0.127$. The red curve represents tight-binding and black line represents DFT bandstructures. The Fermi level is shown with purple dotted line in (\textbf{a-d}). (\textbf{e}) The distance between the nodal points as a function of transverse strain. The solid line represents tight-binding and the symbols represent the DFT results.  }
    \label{fig2}
\end{figure}

At zero transverse strain, i.e. for inter-chain distance $d=3.78\, \ang$ the VB and the CB overlap at the X point of the Brillouin zone by  energy of 14 meV as seen in Figure. \ref{fig2}(a).  If we look at the low energy band dispersion near the Fermi level, it consists of two nodal points separated equally and symmetrically about the X point. It is evident from the Figure. \ref{fig2} (b-d) that there are a couple of two doubly degenerate states emerging, corresponding to each of the two band crossing points, shared by two states from the valence and the conduction bands.
With the increase in transverse compressive strain, the  
CB and the VB move away from each other at the X point (Figure. \ref{fig2}(b-d)) and the metallic phase becomes prominent. Unlike in the 2D or 3D\cite{narang2020topology}, in the 1D Brillouin zone, classification of the material as metallic or semi-metallic is not unambiguous as there can at most be a finite number of states at the Fermi level. Here are only two zero modes shared by conduction and valence electrons. The increase in the distance between the nodal points with transverse strain is indicated in Fig. \ref{fig2}(e). However, from the band-structures obtained from DFT, it was observed that at very large strain, the lowest energy VB and the CBs cross the next higher energy $p_y$ bands (at $d = 3.45 \, \ang$) and with further increase in strain, the  $p_y$ bands form the conduction band minimum and the valence band maximum at the X point (Figure. \ref{fig2}(c), (d)). The linear dispersion and the nodal points exist in the low energy band-structure until the inter-chain separation is $3 \, \ang$ or strain, $\varepsilon_{\mathrm{Tr}}= -0.2$. For inter-chain separation lower than this, the bands arising due to the $p_y$ orbitals cross the Fermi level and the linear dispersion in the low energy bands no longer exists with the nodal points shifting away from the Fermi level. For the range of transverse strain of our interest, i.e. $3.78 \ang > d > 3.45 \ang$, the Fermi velocity $v_F = \frac{t_1 t_2 \sin{\theta}}{\hbar \gamma_1} A$ evaluates to be $\sim 1 \times 10^6 m/s$, which is almost same as that of graphene.

For a host of materials, application of structural deformation has the same effect on the topological properties (e.g. band inversion, semi-metal to topological insulator transition) of the bandstructure as due to the spin-orbit coupling (SOC). Pure carbon materials (graphite or graphene) have low SOC\cite{kane2005quantum} and incorporating its effect in DFT calculation didn't bring any qualitative change in the bandstructure.
In the present work the structural deformation is incorporated by 1) changing stacking i.e., sliding and, 2) applying strain along the length of the coupled chains to check upon the possibility of a gapped phase.

\subsection{Sliding: Transition to the  Semiconductor  Phase}
\begin{figure}[ht]
    \centering
    \includegraphics[width=10cm]{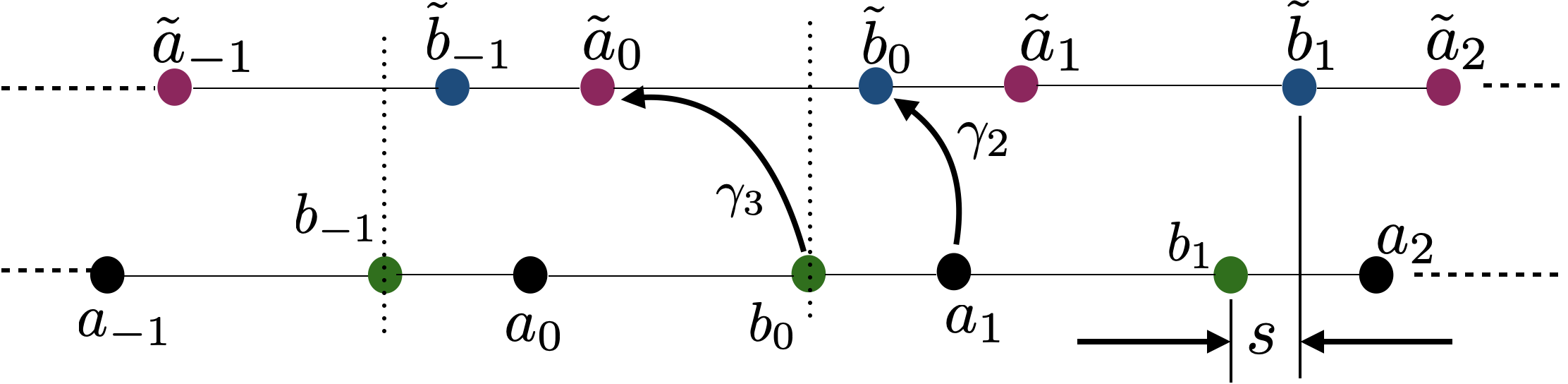}
    \caption{\textbf{Sliding deformation.} Structural deformation of the coupled chain system effected by sliding one chain with respect to another parallelly by distance $s$. The unit cell marked by the vertical dotted lines accordingly gets redefined and the reflection symmetry enjoyed by the structure in AA stacking is broken.}
    \label{slid_chains}
\end{figure}

We now consider the configuration of one of the chains slid with respect to another while keeping them parallel, by a distance $ s$, Fig. \ref{slid_chains}. One can no longer neglect next-to-nearest-neighbour (NNN) inter-chain hopping events. The Hamiltonian for the coupled chains, equation\eqref{double_chain_fourier}, now modified by the effect of sliding $-\delta_2 < s< \delta_2$  should be
\bea \label{double_chain_fourier_slide}
&& H= \sum_{k} \Psi^{\dagger}_k 
\begin{pmatrix} 0 & f_k & \gamma_1  \, e^{i\,k\,s} & \gamma_3  e^{ik (s-\delta_1)}\\ f^{\star}_k & 0 & \gamma_2  e^{ik (s-\delta_2)} & \gamma_1   e^{i\,k\,s}\\ \gamma_1  e^{-i\,k\,s} & \gamma_2  e^{-ik (s-\delta_2)} &0 & f_k \\ \gamma_3  e^{-ik (s-\delta_1)} & \gamma_1  e^{-i\,k\,s}  & f^{\star}_k & 0
\end{pmatrix} \Psi_k. 
\eea
 Note that the characteristic polynomial for the bands of the Hamiltonian in equation\eqref{double_chain_fourier_slide} is a depressed quartic equation\eqref{depressed}, having a non-zero linear term. Hence  evidently the band-structure does not have symmetry about the zero-energy level, breaking the particle-hole symmetry and hence getting out of BDI topological class. Being a 1D system, thus \eqref{double_chain_fourier_slide} falls in the topologically trivial class. 
 
 For small slide $s \ll \delta_2$, $\gamma_2 \sim \gamma_3 \ll t_1$ is a good approximation \footnote{In the figure \ref{slid_chains}, the difference between $\delta_1$ and $\delta_2$ has been exaggerated. But in reality, $\delta_2 \sim 0.97 \delta_1$  and hence according to the parametrization later introduced in equation\eqref{gap_param} this approximation is well justified.} and solving perturbatively at the node $k_1$, the analysis in the Appendix shows emergence of a finite bandgap in equation\eqref{gap_pert}, linearly proportional to $\gamma_2$:
\bea \label{pert_gap1}
\Delta = \gamma_2 \frac{(|t_1 -t_2| ) \left((t_1+t_2)^2 - \gamma^2_1 \right)}{2 \gamma_1 t_1 t_2}+\mathcal{O}(\gamma^2_2).
\eea
This clearly indicates a transition from the metallic to the semiconducting phase. As is clear from Figure \ref{slid_chains}, sliding deformation breaks reflection symmetry of the stacking pattern. The degeneracy of the two states belonging to conduction and valence  bands occurring at each of the nodal points gets lifted, opening up a bandgap. However to determine the nature of the bands close to the nodal points and to compare with  DFT results, we first model the slide dependence of tight-binding parameters as follows and diagonalize the Hamiltonian in equation\eqref{double_chain_fourier_slide} numerically. 
\begin{figure}[ht]
    \centering
    \includegraphics[width=\columnwidth]{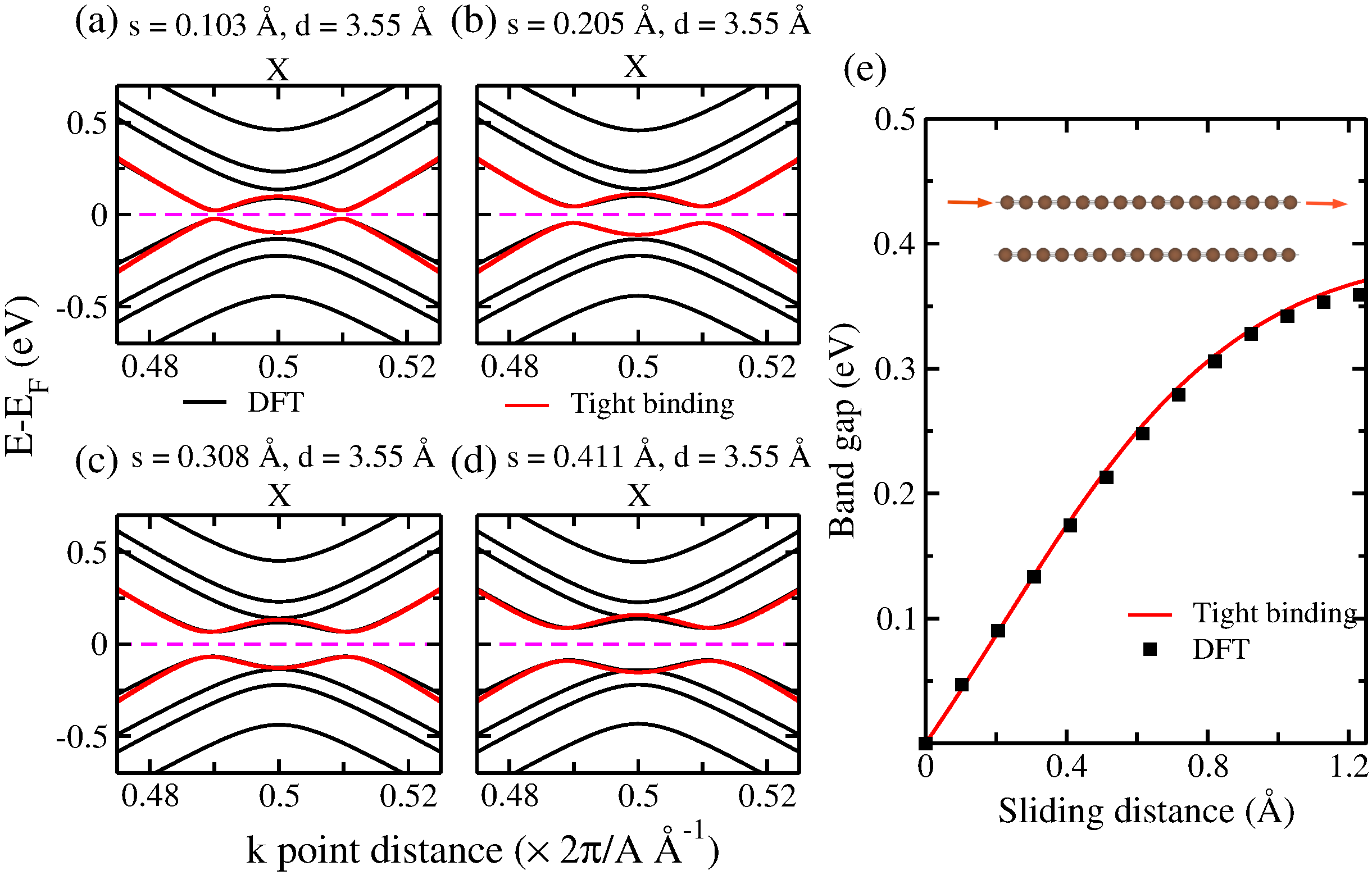}
    \caption{\textbf{Effect of sliding on bandstructure - gap opening.} 
    The chains are subjected to a transverse strain of $\varepsilon_{\mathrm{Tr}} = -0.061$ so that the inter-chain distance is $3.55 \, \ang$. Panels (\textbf{a})-(\textbf{d}) depicts larger bandgap opening between the CB and the VB near Fermi level with increasing slide parameters $s = 0.103 \, \ang, 0.205 \, \ang, 0.308 \, \ang$ and $0.411 \, \ang$ respectively. The red and the black lines respectively are from Hubbard model tight-binding and DFT calculations, showing quite satisfactory matching. Panel (\textbf{e}) shows the value of the bandgap as a function of slide distance with high degree of agreement between tight-binding and DFT simulation for slide amount up to $ 1.2\, \ang (\sim \delta_2)$.}
    \label{slide_bstr}
\end{figure}
We first observe that not only the NNN parameters $\gamma_2, \gamma_3$ depend upon slide distance $s$, its dependence of $\gamma_1$ should be incorporated as well in the Hamiltonian in equation\eqref{double_chain_fourier_slide}. The later can be introduced in a straightforward manner starting from equation\eqref{par}
\bea \label{ds}
\gamma_1 (d,s) = t_0 \exp \left(- \frac{s^2 + d^2}{\kappa \, d^2_0}\right).
\eea
For modelling the other two parameters $\gamma_2, \gamma_3$, we keep in mind that the Hamiltonian in equation\eqref{double_chain_fourier_slide} should match  the pure AA configuration Hamiltonian, equation\eqref{double_chain_fourier} as the parameter $s \rightarrow 0$. This results into the following:
\bea \label{gap_param}
&& \gamma_2 (s) = \Tilde{t}_0 \left(\exp \left[-\frac{ \left(s^2-2 \delta_2 \,|s|  \right)}{ \delta^2_2}\right] - 1\right)\nonumber \\&& \gamma_3 (s) = \Tilde{t}_0 \left(\exp\left[-\frac{ \left(s^2-2 \delta_1 \,|s|  \right)}{ \delta^2_1}\right] - 1\right)
\eea
where $\Tilde{t}_0 = 0.183 \, eV$ is the interaction strength. In contrast to equation\eqref{ds}, the gaussian width control parameter in equation\eqref{gap_param} has been chosen to be 1. The above model of the parameters in equation\eqref{gap_param} obviously is not periodic over the lattice and is valid up to good numerical agreement with bandgap results from DFT for slide $ -\delta_2< s < \delta_2$. The band dispersion for the lowest energy bands, calculated as per the above tight-binding formulation is plotted together with the respective DFT bandstructures in Figure \ref{slide_bstr} (a-d) for various sliding distances. We have kept the inter-chain distance as $d=3.55 \, \ang$ for all the slid structures. This value $d$ was chosen because the lowest energy bands near the Fermi level do not touch the next higher $p_y$ bands and the nodal points are well separated in the bandstructure for zero-slide as shown in Figure \ref{fig2} (b). The lifting of degeneracy at the nodal point and the emergence of a bandgap while turning on sliding is clearly exemplified in the Fig. \ref{slide_bstr}. 
The lifting of the degeneracy and opening of bandgap starting from a metallic phase is reminiscent of the phenomenon of band-inversion associated with topological phase transition in the edge states in higher dimensional systems\cite{bansil2016colloquium}, with strong SOC.

A plot of bandgap $vs.$ sliding distance, as shown in Figure \ref{slide_bstr} (e) shows a linear increase in bandgap for low sliding, reaching maximum ($\sim$ 0.36 eV) at a sliding distance equal to half of the lattice parameter and then decreases symmetrically for further sliding. The bandgap increases linearly for small slide up to: $s \sim 0.4 \ang$  (Fig. \ref{slide_bstr} (e)). It is evident from the structure that the bandgap is an even function of the slide parameter making it a non-smooth function  at $s=0$, a routine behaviour for topological insulators with tunable bandgap\cite{scharf2019tuning}.

\subsection{Longitudinal strain}
\begin{figure}[ht]
    \centering
    \includegraphics[width=\columnwidth]{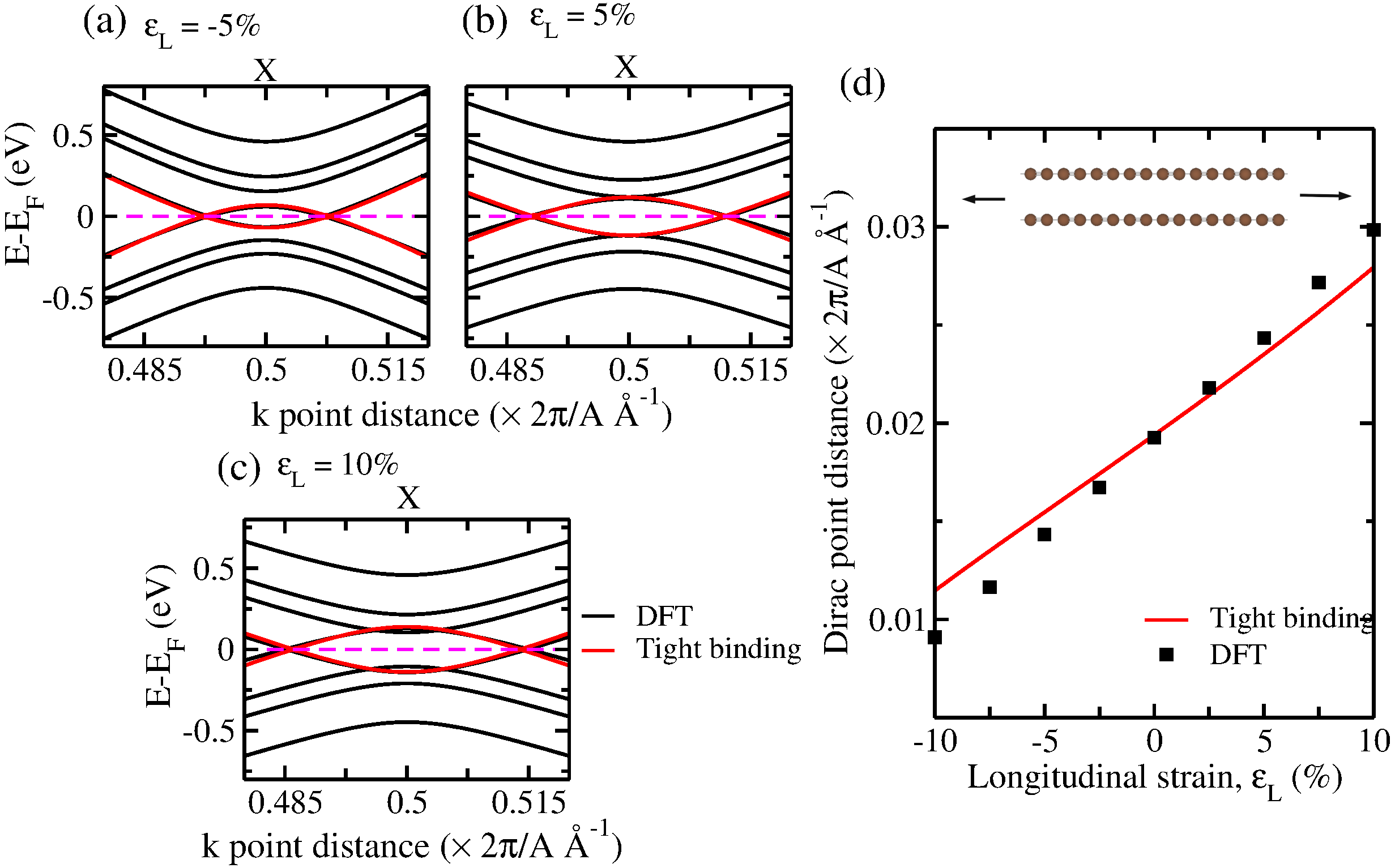}
    \caption{\textbf{Effect of compressive and tensile longitudinal strain on bandstructures.} With higher compression the nodal points come closer to each other. Panels (\textbf{a-c}) represent longitudinal strain $\varepsilon_{\mathrm{L}} = -5\%, +5\%$ and $10\%$, respectively. Here negative strain denotes compression. The bands nearest to the Fermi level start touching the higher band for compression more than $5\%$. Panel (\textbf{d}) shows a plot of separation between the nodal points as a function of strain percentage.}
    \label{LongStr_bstr}
\end{figure}
 Maintaining pure AA stacking of the coupled polyyne chains and application of strain along the length of it, either as compression or elongation changes the bond lengths $\delta_{1,2}$ by a fraction. The effect of longitudinal strain on the bandstructure from equation\eqref{2_chain_BS} can be incorporated by tuning the parameters $t_1$ and $t_2$ and by bringing in the fractional change again via the gaussian dependence on the fraction $r$  of compression:
\bea
t_1 (r) = t_1 \exp(1- r^2), ~ t_2 (r) = t_2 \exp(1- r^2).
\eea
The strain fraction, $r$ is related to the percentage of the longitudinal strain as $ \varepsilon_{\mathrm{L}} = 100 (1-r) \%$. For the range of separation between the two chains, where two distinct nodal points are observed (ie. roughly from 3.55 \AA \; to 3.8 \AA), the distance between the two nodes depends on the strain fraction as:
\bea \label{strain_dist}\left(2 \pi - 2 \arccos \left(\frac{\gamma^2_1 - t^2_1 (r) - t^2_2 (r)}{2\,t_1 (r)\,t_2 (r)} \right) \right)/A,
\eea as already derived earlier. 

The lowest energy bands, nearest to the Fermi level, are plotted in Figure. \ref{LongStr_bstr} (a,b,c) for varying longitudinal strain. The effect of strain on the bands is clearly visible in these plots, where an increased tensile strain increases in the separation between the nodal points. The separation between the nodal point is plotted in Figure. \ref{LongStr_bstr} (d) as function of longitudinal strain percentage for both DFT and tight-binding calculations. The plot shows a linear dependence on the strain and for lower strain values both the methods show good agreement.

\section{Conclusion}

In this work, we studied several interesting aspects of electronic bandstructures of a system of coupled parallel polyyne chains. While a single polyyne chain is insulating with a finite bandgap, the coupled system even when kept at a relaxed separation, has states at the Fermi level and the system is conducting. When the chains are brought closer, the VB and the CBs cross showing metallic behaviour. The exact locations of the nodal points in the Brillouin zone and the Fermi velocity near the Fermi level have been corroborated by a first-principles DFT calculation. The application of longitudinal strain also preserves the nodal points. This metallic phase is robust under structural perturbations like these strains.

As a dramatic consequence of the introduction of further structural deformation of sliding, a bandgap opens up. The gap is tunable and increases linearly for small slide. We have predicted the gap energy again by simulation as well as analytical calculations. In the tight-binding picture, the transition from the metallic to the semiconducting phase is a transition from a topologically non-trivial BDI phase to a trivial phase.

Apart from experimental realization, either with real polyyne chains or quantum simulations of those, there are some open questions that warrant further investigations, such as the calculation of the topological index in the metallic phase, approximately modelling the semiconducting phase as a topologically non-trivial phase and hence the classification of the corresponding quantum phase transition.

\section*{Acknowledgements}
RB thanks the Fulbright Foundation and the SERB Govt of India (SRG/2020/001037 and CRG/2020/002035) for support.
SB would like to acknowledge SERB, Govt. of India (SRG/2020/000562 and CRG/2020/000434) and BITS Pilani K. K. Birla Goa Campus, India (GOA/ACG/2019-20/NOV/08) for the financial support.
Correspondence with Martin Zirnbauer and discussions with Grigory Tarnopolsky and Indrakshi Raychowdhury are thankfully acknowledged.


\section*{Competing interests}
The authors declare no competing interests.


\appendix
\section*{Appendix: Perturbative analysis of bandgap in sliding} \label{appanalytical}
For small amount of slide from AA configuration, and with the well justified assumptions that $\gamma_2 \sim \gamma_3 \ll \gamma_1$, we can try to probe the bands near Fermi level, treating $\gamma_2$ as a perturbation parameter. The characteristic polynomial of the kernel of the Hamiltonian in equation\eqref{double_chain_fourier_slide} is given by the following depressed quartic equation in $\lambda$:
\bea \label{depressed}
&&\lambda^4 + (C_0 (k)-2\gamma^2_2)\lambda^2 + \gamma_2 D_0(k) \lambda + (E_0(k) + \gamma^2_2 E_1 (k) + \gamma^4_2) =0.
\eea
Here the coefficients are
\bea \label{coefficeients} && C_0 (k) = -2 (\gamma^2_1  + t^2_1 + t^2_2 + 2 t_1 t_2 \cos(kA)), ~ D_0 (k) = -8 \gamma_1 (t_1 + t_2)  \cos^2(kA/2) , \nonumber \\
&& E_0 (k) = \alpha + \beta_1 \cos(kA) + \beta_2 \cos(2kA), E_1 (k) = -  (\Tilde{\alpha} + \Tilde{\beta} \cos(kA)), \nonumber
\eea
where,
\bea
&&\alpha = \left( \gamma^2_1 - t^2_1 - t^2_2\right)^2 + 2 t^2_1 t^2_2, ~ \beta_1 = 4 t_1 t_2 \left( t_1^2 + t^2_2 \right) - 4 \gamma^2_1 t_1 t_2 \nonumber\\
&&\beta_2 = 2t^2_1 t^2_2 ,\, \Tilde{\alpha} = 4 t_1 t_2, \Tilde{\beta} = 2 \gamma^2_1 + 2 (t^2_1+t^2_2).
\eea
We seek to solve it perturbatively at the nodal point $ k_1 = \arccos\left(\dfrac{\gamma^2_1 -t^2_1- t^2_2}{2 t_1 t_2}\right)/A$.

Particularly focusing at the point $k_1$, we start with a perturbative expansion $\lambda = \lambda_0 + \gamma_2 \lambda_1 + \gamma^2_2 \lambda_2 + \cdots$. Now, for the two bands, which are closest to the Fermi level, $\lambda_0 = 0$ at $k=k_1$ first order perturbation yields:
\bea \label{pert_gap}
\lambda^{\pm}_1 = \dfrac{-D_0 (k_1) \pm \sqrt{D^2_0 (k_1) - 4C_0 (k_1) E_1 (k_1)}}{2C_0(k_1)}.
\eea
From equation\eqref{pert_gap} we readily infer that the non-zero bandgap, that has resulted due to small slide is given by
\bea \label{gap_pert}
\Delta = \gamma_2  |\lambda^+_1 - \lambda^-_1|  + \mathcal{O}(\gamma^2_2) &=& \gamma_2 \Big{|}\dfrac{ \sqrt{D^2_0 (k_1) - 4C_0 (k_1) E_1 (k_1)}}{C_0 (k_1)}\Big{|}+ \mathcal{O}(\gamma^2_2)\nonumber \\
&=& \gamma_2 \frac{(|t_1 -t_2| ) \left((t_1+t_2)^2 - \gamma^2_1 \right)}{2 \gamma_1 t_1 t_2}+\mathcal{O}(\gamma^2_2).
\eea


\end{document}